\documentclass[%
 reprint,
superscriptaddress,
 amsmath,amssymb,
 aps,
floatfix,
]{revtex4-1}

\usepackage{graphicx}
\usepackage{dcolumn}
\usepackage{bm}

\begin{document}

\title{Modeling temporal networks with bursty activity patterns of nodes and links}

\author{Takayuki Hiraoka}
\affiliation{Asia Pacific Center for Theoretical Physics, Pohang 37673, Republic of Korea}
\author{Naoki Masuda}
\affiliation{Department of Mathematics, University at Buffalo, State University of New York, Buffalo, NY 14260-2900, USA}
\affiliation{Computational and Data-Enabled Science and Engineering Program, University at Buffalo, State University of New York, Buffalo, NY 14260-5030, USA}
\author{Aming Li}
\affiliation{Department of Zoology, University of Oxford, Oxford OX1 3PS, United Kingdom}
\affiliation{Department of Biochemistry, University of Oxford, Oxford OX1 3QU, United Kingdom}
\author{Hang-Hyun Jo}
\email{hang-hyun.jo@apctp.org}
\affiliation{Asia Pacific Center for Theoretical Physics, Pohang 37673, Republic of Korea}
\affiliation{Department of Physics, Pohang University of Science and Technology, Pohang 37673, Republic of Korea}
\affiliation{Department of Computer Science, Aalto University, Espoo FI-00076, Finland}

\date{\today}

\begin{abstract}
The concept of temporal networks provides a framework to understand how the interaction between system components changes over time. In empirical communication data, we often detect non-Poissonian, so-called bursty behavior in the activity of nodes as well as in the interaction between nodes. 
However, such reconciliation between node burstiness and link burstiness cannot be explained if the interaction processes on different links are independent of each other. This is because the activity of a node is the superposition of the interaction processes on the links incident to the node and the superposition of independent bursty point processes is not bursty in general.
Here we introduce a temporal network model based on bursty node activation and show that it leads to heavy-tailed inter-event time distributions for both node dynamics and link dynamics. Our analysis indicates that activation processes intrinsic to nodes give rise to dynamical correlations across links. Our framework offers a way to model competition and correlation between links, which is key to understanding dynamical processes in various systems. 
\end{abstract}

\maketitle


\section{Introduction}

Temporal networks have become an important framework to understand the dynamics of complex systems over the last decade~\cite{Holme2012Temporal, Masuda2016Guide, Holme2019Temporal}. By integrating the topological knowledge of a system, described by a graph, with the information about the temporal nature of the interaction between its components, represented by time series, we can precisely track who interacts with whom and when. The interaction dynamics can be captured at several different levels. First, the interaction between each pair of nodes specifies the dynamics of the link. Second, by aggregating the interaction between a node and all of its neighbors, one obtains the dynamics of the node, which shows how the node interacts with others. Lastly, by collecting all the interaction between every pair of nodes, one can tell about the dynamics of the entire system. For instance, in communications systems where people interact by sending messages, the link dynamics corresponds to the message correspondence between a pair of individuals, while the node dynamics corresponds to the inbox of messages sent or received by an individual.

Human communication is known to exhibit non-Markovian, inhomogeneous temporal patterns which are commonly referred to as being \textit{bursty}~\cite{Barabasi2005Origin, Karsai2018Bursty}. When each communication event is instantaneous or lasts for a short period so that its duration can be neglected compared to other time scales, one can regard the communication sequence as a realization of a point process. The burstiness of a point process is mainly characterized by a heavy-tailed distribution of time intervals between consecutive events, or inter-event times (IETs), in contrast to Poisson processes for which the IET distributions are exponential.

Interestingly, empirical data suggest that in communications systems, the communication sequences of nodes and of links are both characterized by power-law distributions with a similar scaling exponent~\cite{Karsai2012Correlated, Saramaki2015Seconds}. This cannot be taken for granted for the following reason. As mentioned above, the communication sequence of a node is the superposition of the communication events on all the links between the node and its neighbors. However, a superposition of independent renewal processes does not retain the statistics of the original processes in general. 
In fact, the IET distribution for the superposed process tends to an exponential distribution in the limiting case where the number of independent source processes is large~\cite{Cox1954Superposition, Hoopen1966Superposition, Lindner2006Superposition}. Therefore, the observation that both node dynamics and link dynamics are bursty suggests the presence of correlations across communication processes on different links. Such link-link correlations can have a significant impact on the dynamical processes taking place in the network~\cite{Miritello2011Dynamical, Kivela2012Multiscale, Backlund2014Effects, Saramaki2015Seconds}, but their origin has yet to be understood.

Here, we study the mechanisms behind the burstiness in node and link activity patterns by considering a model in which the nodes are activated randomly in time with non-Poissonian statistics and two nodes may communicate if and only if they are simultaneously activated. In Sec.~\ref{sec:model}, we introduce two variants of the model with different communication rules. In Sec.~\ref{sec:numerical}, we report results of numerical simulations performed on networks with various topologies. We show that, for both models and for all the networks, the communication patterns are characterized by heavy-tailed IET distributions for both nodes and links. Section~\ref{sec:analysis} is devoted to explaining the origin of the burstiness in node and link activity patterns for each model. We describe the behavior of the model in a system of two nodes by relating it to the statistics of the sum of a random number of random variables. We use the same approach to derive the activity patterns of nodes and links in larger networks. Finally, we conclude our work in Sec.~\ref{sec:conclusion}.

\section{Model}
\label{sec:model}

We consider a network of size $N$ with a given structure, in which each node is activated randomly at discrete times and its activation pattern is modeled by a renewal process with a given inter-activation time (IAT) distribution, denoted by $P(r)$. In order to start the activation process at equilibrium, the first activation time $t_0\geq 0$ of each node is assigned according to the residual time distribution~\cite{Cox1962Renewal} 
\begin{equation}
    P_0(t_0)=\frac{1}{\langle r \rangle}\sum_{r = t_0}^\infty P(r),
    \label{eq:residual_distribution}
\end{equation}
where $\langle r \rangle$ denotes the average IAT. The node is then activated at times $t_l = t_0 + \sum_{l'=1}^l r_{l'}$ for $l = 1, 2, \cdots$, where each IAT, denoted by $r$, is independently drawn from $P(r)$ \footnote{The dummy variable $l'$ is omitted for the sake of simplicity. The same notation rule applies in the rest of the paper.}. In our work, we adopt a power-law IAT distribution,
\begin{equation}
    P(r)=\frac{r^{-\alpha}}{\zeta(\alpha)}\ \textrm{for}\ r=1,2,\cdots,
    \label{eq:power_law_IAT_distribution}
\end{equation}
where $\zeta(\alpha) \equiv \sum_{x = 1}^\infty 1 / x^\alpha$ is the Riemann zeta function. We choose $\alpha > 2$ to make Eq.~\eqref{eq:residual_distribution} converge.

\begin{figure}[tb]
    \centering
    \includegraphics[width=0.95\columnwidth]{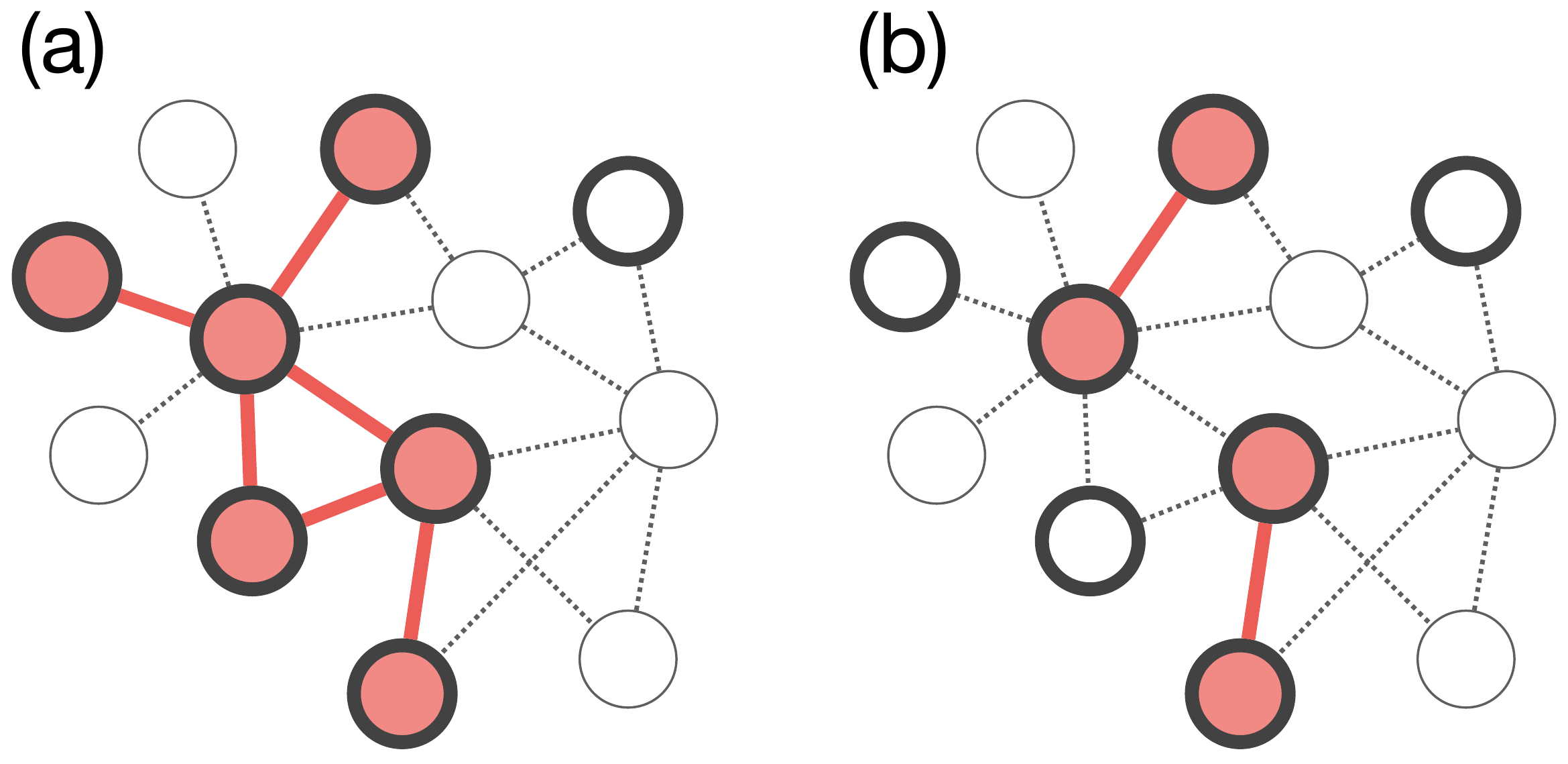}
    \caption{Schematically illustrated snapshots of communication according to the two models given the same set of activated nodes, which are enclosed by thick solid lines. The red filled circles and red solid lines represent the nodes and links with a communication event, respectively.
    (a) The polyvalent model assumes that the activated nodes communicate with all the neighbors that are simultaneously activated. 
    (b) In the monovalent model, each activated node communicates with at most one neighbor. }
    \label{fig:schematic}
\end{figure}

At each time step $t$ ($0 \leq t \leq T$), pairs of activated nodes communicate with each other. As depicted in Fig.~\ref{fig:schematic}, here we consider two variants of the model.
The first variant, which we call the \textit{polyvalent} model, assumes that an activated node communicates with all the activated neighbors. The case where an activation does not lead to communication only occurs when the node does not have any simultaneously activated neighbors. 
In the second variant, which we refer to as the \textit{monovalent} model, an activated node is randomly paired with one of its activated neighbors to have at most one communication partner at the same time, as is the case for one-to-one phone calls. An activated node cannot communicate with others if none of the neighbors are simultaneously activated, or if all the simultaneously activated neighbors are already paired with other nodes.

In both models, the communication events between a pair of adjacent nodes can be associated not only with the link but also with the nodes. In other words, we can define a communication event sequence for each link as well as for each node, the latter being the union of the communication events on all the links attached to the node. Hereafter, we refrain from the wording ``inter-event time" to avoid confusion and instead use ``inter-communication time" (ICT) to represent the time interval between consecutive communication events on the sequence affiliated with a node or with a link. Note that the communication sequence of a node does not agree with the activation sequence of the node in general, because an activated node may not communicate with anyone, as shown in Fig.~\ref{fig:schematic}. We denote the ICTs by $\tau$. Whenever necessary, subscripts $i$ and $ij$ will distinguish between node $i$'s properties and link $ij$'s properties; sub- or superscripts $\mathrm{p}$ and $\mathrm{m}$ will indicate variables and functions related to the polyvalent and monovalent models, respectively. In the following sections, we discuss the statistics of node and link ICTs.

\section{Numerical Results}
\label{sec:numerical}
\begin{figure*}[tb]
    \centering
    \includegraphics[width=\textwidth]{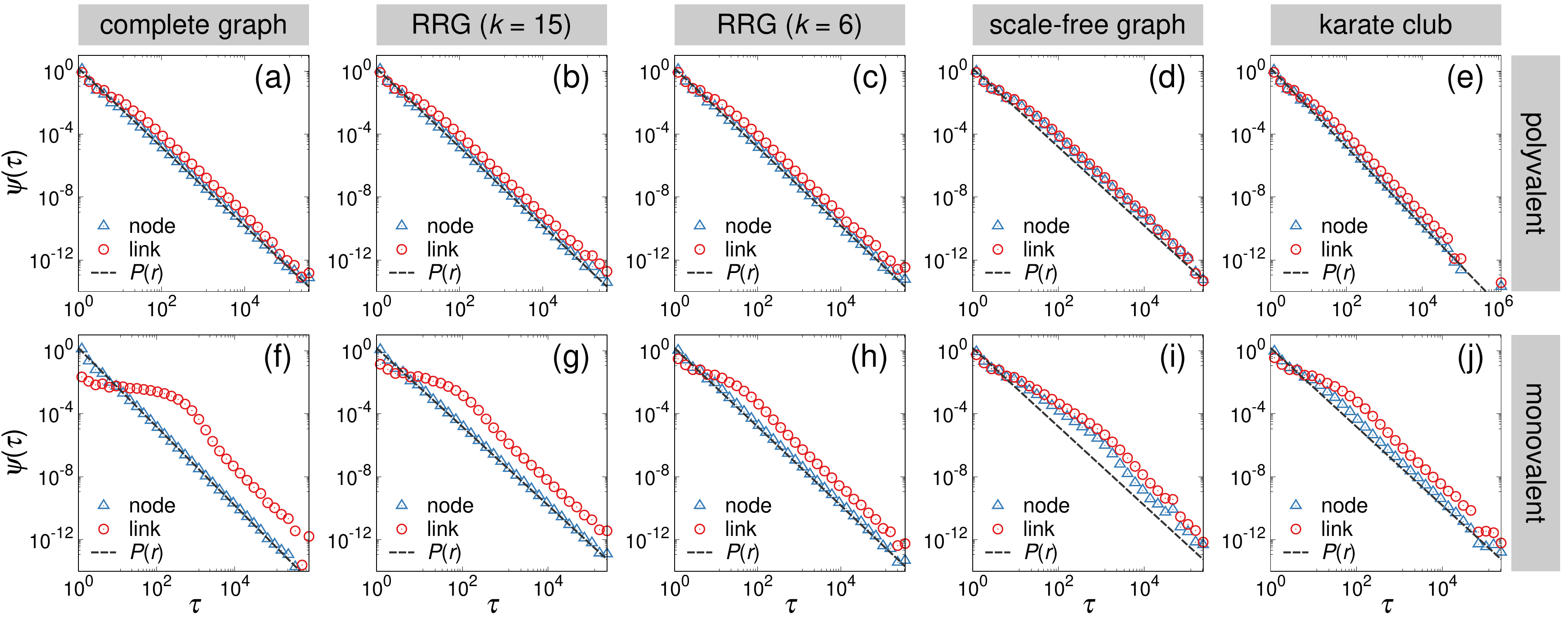}
    \caption{The inter-communication time (ICT) distributions $\psi(\tau)$ of nodes and links for the polyvalent model (top panels) and for the monovalent model (bottom panels). The network structures are, from left to right, a complete graph (a,~f), random regular graphs with degree $k = 15$ (b,~g) and with $k = 6$ (c,~h), a scale-free graph with degree distribution $\propto k^{-2.1}$
    (d,~i), and Zachary's karate club network (e,~j). The parameters used are $N = 100$ and $T = 10^7$ for the complete and random regular graphs, $N = 1000$ and $T = 10^6$ for the scale-free graph, and $N = 34$ and $T = 10^7$ for Zachary's karate club network. The inter-activation time (IAT) distribution, which follows Eq.~\eqref{eq:power_law_IAT_distribution} with $\alpha = 2.5$, is represented by the dashed line.}
    \label{fig:ICT_distributions}
\end{figure*}

We carry out numerical simulations for synthetic networks with different topologies such as complete graphs, random regular graphs, and scale-free graphs, as well as Zachary's karate club network~\cite{Zachary1977Information} as an example of a real-world network. Figure~\ref{fig:ICT_distributions} summarizes the node and link ICT distributions $\psi(\tau)$. Here we set $\alpha = 2.5$.

The polyvalent model yields the node and link ICTs both of which are distributed almost indistinguishably from the power-law IAT distributions for the various network topologies. In contrast, the monovalent model results in different communication patterns depending on the network structure. For homogeneous graphs such as complete and random regular graphs, the node ICT distributions are almost identical to those of IATs, while the link ICT distributions show a hump at short time scales that is not in the power-law IAT distributions. As we make the network sparser by reducing the degree of random regular graphs, the hump becomes smaller and the range of $\tau$ in which the distribution approximately follows a power law becomes wider, implying that the sparseness of networks plays an important role in realizing bursty communication patterns on links.

\begin{figure}[tb]
    \centering
    \includegraphics[width=\columnwidth]{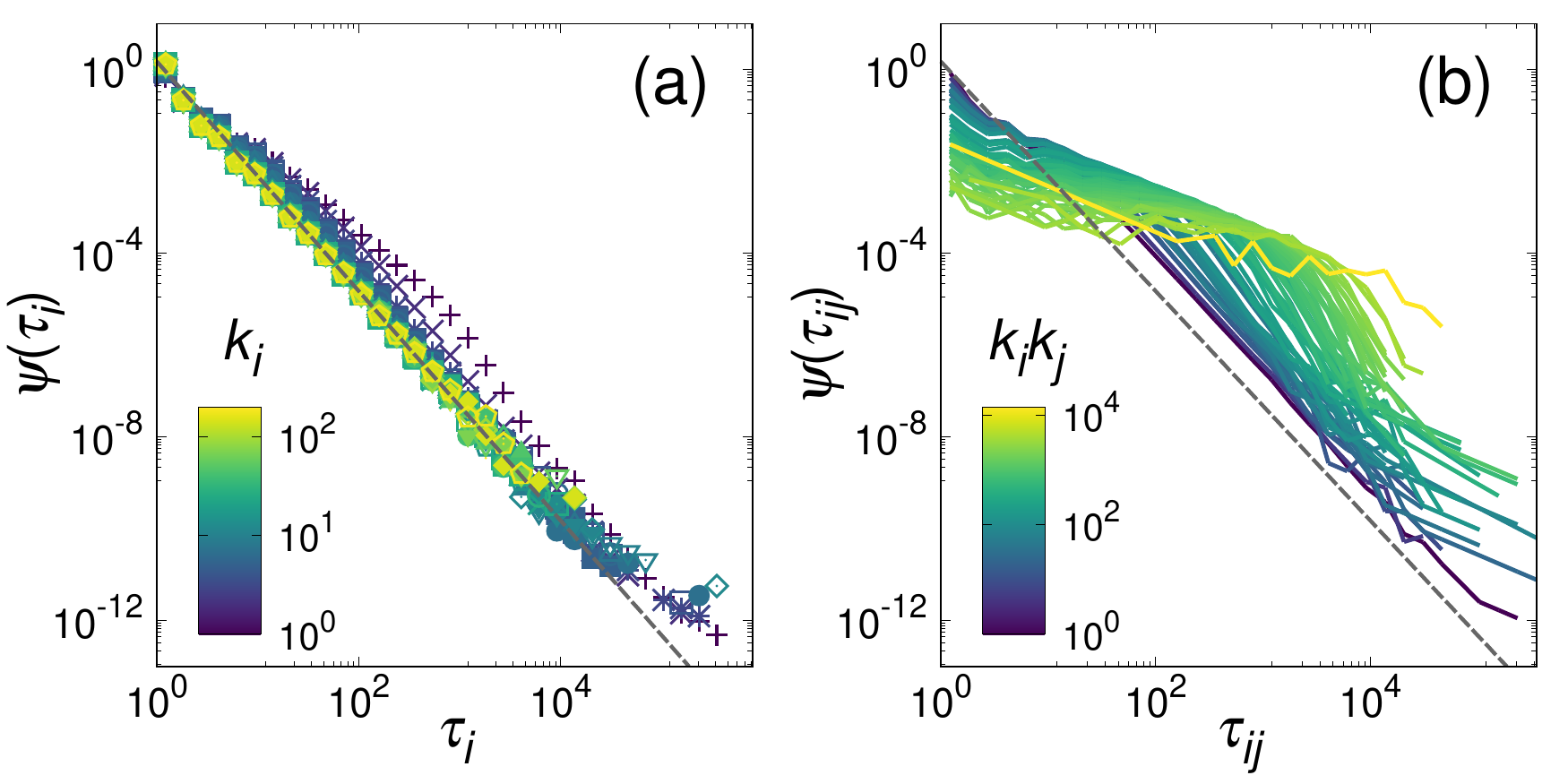}
    \caption{(a) Node ICT distributions grouped by degree $k_i$. (b) Link ICT distributions grouped by the product of the degrees of the end nodes, $k_i k_j$. Both figure panels are based on the simulation results for the monovalent model on the scale-free graph shown in Fig.~\ref{fig:ICT_distributions}(i).
    }
    \label{fig:degree_based_ICT_distributions}
\end{figure}

For scale-free graphs, both the node and link ICTs differ from the IATs in terms of the distribution. To examine the effect of structural heterogeneity, we group the nodes by degree and consider the node ICT distribution for each group. Figure~\ref{fig:degree_based_ICT_distributions}(a) shows that the deviation between the ICT and IAT distributions is larger for nodes with smaller degrees. This can be understood intuitively as follows: When activated, a node with more neighbors is more likely to find a communication partner, i.e., a neighbor who is simultaneously activated and available. In the extreme case where the degree of a node is large enough so that the node almost always finds a partner whenever activated, the node ICT distribution would coincide with the IAT distribution. On the other hand, the ICT distribution would deviate more from the IAT distribution for nodes with less neighbors because of the difficulty of finding a partner.

Similarly, we group the links by the product of the degrees of the end nodes of the link, $k_i$ and $k_j$, and measure the distribution of $\tau_{ij}$ conditioned on each value of $k_i k_j$. As shown in Fig.~\ref{fig:degree_based_ICT_distributions}(b), the ICT distributions for links with larger values of $k_i k_j$ show larger deviations from the IAT distribution. Suppose that two adjacent nodes are simultaneously activated. If the degrees of the nodes are larger, then the nodes are likely to have larger numbers of other simultaneously activated neighbors that they potentially communicate with. Therefore, the probability that the two nodes communicate with each other decreases. In contrast, if the degrees are small, then the two nodes are more likely to communicate with each other. We discuss these intuitions more quantitatively in the following Section.

Finally, for Zachary's karate club network, the results are similar to those for the homogeneous graphs. The node ICT distribution shows a small deviation from the IAT distribution, which is due to the degree heterogeneity.

\section{Analytical results}
\label{sec:analysis}

In this section, we provide an analytical examination of the behavior of the models. For that, we start by considering a minimal system that consists of a pair of adjacent nodes, which we call a \textit{dimer}. We show that the number of IATs that compose each ICT is a random variable that follows a power-law distribution. Thus, we can describe an ICT as the sum of a random number of random variables, and we show that the sum is also distributed as a power law. Then, we derive the statistics of the link and node ICTs for the polyvalent model directly from the results for a dimer. Finally, we describe the monovalent communication as a result of random success and failure of the polyvalent communication. This leads to expressions of link and node ICTs for the monovalent model as a geometric sum of polyvalent link and node ICTs, respectively. We obtain power-law statistics in this case as well.

\subsection{Case of dimers}
\label{subsec:dimer}
A dimer is a pair of nodes connected only to each other. Under this setup, the monovalent model is equivalent to the polyvalent model because the nodes have no other nodes to communicate with except for each other. Moreover, the communication sequences of the two nodes are identical to each other as well as to that of the link connecting them.

\begin{figure*}[tb]
    \centering
    \includegraphics[width=\textwidth]{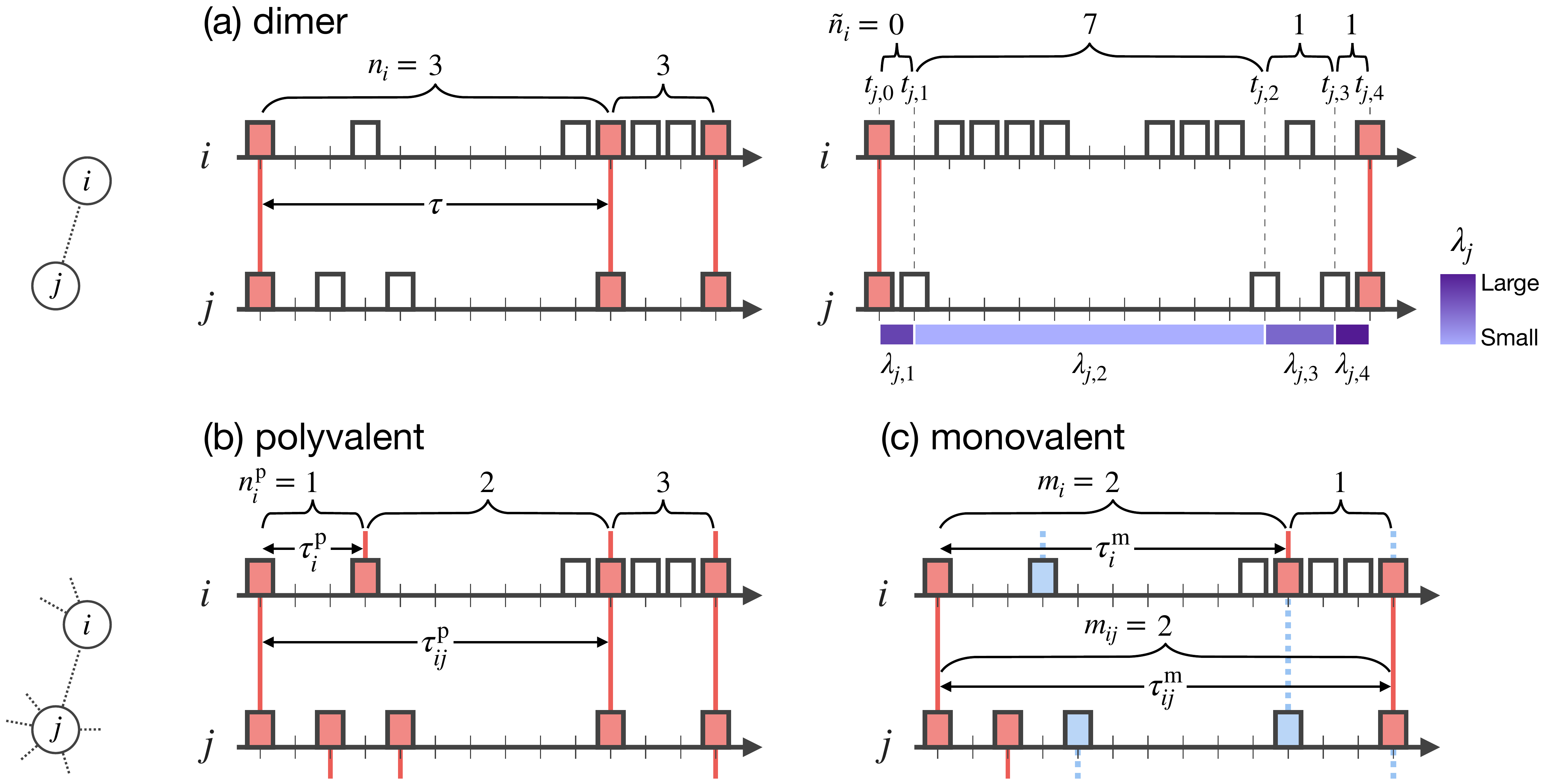}
    \caption{
    Schematic illustrations of communication sequences between a pair of nodes $i$ and $j$. The red and white rectangles represent activations with and without communication, respectively. The solid vertical lines identify communication events between the pair. In (b) and (c), nodes are activated with the same temporal pattern as in the left panel of (a). 
    (a) The case where the two nodes form a dimer. Variable $n_i$ denotes the number of activations (represented by rectangles) of node $i$ between two communication events (red rectangles).
    The right panel in (a) shows how the activation pattern of node $j$ can be mapped to an activation-modulated Poisson process where each IAT is associated with an activation probability $\lambda_j$. Variable $\tilde{n}_i$ denotes the number of activations of $i$ in each period segmented by activations of $j$. 
    (b) The polyvalent model. The communication pattern on the link is the same as that of the dimer, while each node also communicates with other neighbors and experiences more frequent communication with smaller ICANs and ICTs, which are denoted by $n^\mathrm{p}_i$ and $\tau^\mathrm{p}_i$, respectively. 
    (c) The monovalent model. Simultaneous activation of adjacent nodes may fail to trigger communication, as indicated by light blue rectangles and vertical dotted lines. 
    Variable $m_{ij}$ denotes the number of events that nodes $i$ and $j$ are simultaneously activated (represented by vertical lines) between two consecutive communication events on link $ij$ (solid vertical lines),  while $m_i$ denotes the number of activations of $i$ that concurred with activations of any of its neighbors (colored rectangles) between two consecutive communication events of $i$ with one of its neighbors (red rectangles). See main text for details.
    }
    \label{fig:schematic2}
\end{figure*}

In general, each of the two nodes of the dimer, denoted by $i$ and $j$, can be activated more than once between two consecutive communication events, as sketched in Fig.~\ref{fig:schematic2}(a). This leads to expressions of an ICT, denoted by $\tau$, as the sum of successive IATs of each node:
\begin{equation}
    \tau = \sum_{n'=1}^{n_i} r_{i, n'} = \sum_{n'=1}^{n_j} r_{j, n'},
    \label{eq:ICT_is_equal_to_sum_of_IAT}
\end{equation}
where $r_{\omega, n'}$ denotes the $n'$th IAT of node $\omega \, (\omega = i, j)$ within the ICT and $n_\omega$ denotes the number of times that node $\omega$ is activated between the two communication events. We call $n_\omega$ an inter-communication activation number (ICAN) of node $\omega$. The random variables $n_i$ and $n_j$ will have the same statistics by symmetry, which allows us to focus on node $i$'s point of view from now on. Keeping Eq.~\eqref{eq:ICT_is_equal_to_sum_of_IAT} in mind, our goal is (i) to derive the statistics of $n_i$ and (ii) to compute the statistics of $\tau$ as the sum of the $n_i$ independent random variables $r_i$.

Let us consider the activation processes of the two nodes between two consecutive communication events as follows (see the right panel of Fig.~\ref{fig:schematic2}(a)). Node $j$ is activated and communicates with node $i$ at time $t_{j, 0}$ for the first time, and then activated at $t_{j, 1}, t_{j, 2}, \dots, t_{j, n_j - 1}$ until it communicates with $i$ again at $t_{j, n_j}$ for the second time. The number of activations of node $i$ between the two consecutive activations of node $j$ at time $t_{j, n'-1}$ and $t_{j, n'}$ is denoted by $\tilde{n}_{i, n'}$, where $1 \leq n' \leq n_j$. The ICAN is then written as the sum of the numbers of activations in each segment indexed by $n'$,
\begin{equation}
    n_i = \sum_{n'=1}^{n_j}\tilde{n}_{i, n'}.
\end{equation}

In order to derive the distribution of ICANs, we map the renewal process of the activations of node $j$ to an inhomogeneous Bernoulli process, in which the activation probability is a time-dependent parameter. Particularly, we adopt the framework of mapping a continuous renewal process into an event-modulated Poisson process~\cite{Masuda2018Gillespie}. An event modulated Poisson process is a process where the event rate $\lambda$ is independently redrawn from a distribution $F(\lambda)$ after every event and remains constant until the next event occurs with that rate. The cumulative IET distribution is then shown to be the Laplace transform of $F(\lambda)$~\cite{Masuda2018Gillespie}.

In our case, we should instead consider an activation-modulated Bernoulli process since time is discrete. In this framework, node $j$ is activated at each time step with a probability $\lambda_j$, which is independently redrawn from a distribution $F(\lambda_j)$ upon every activation. Then, from node $i$'s point of view, each of its $\tilde{n}_{i, n'}$ activations between two consecutive activations of node $j$ can be considered as an independent Bernoulli trial with the success probability $\lambda_{j, n'}$ that node $j$ is activated at the same time (see the right panel of Fig.~\ref{fig:schematic2}(a)). 

Now, we hypothesize that a large $n_i$ is likely to be dominated by the numbers of activations that occur within a few long IATs of node $j$ governed by small activation rates. This leads to a simplification of the argument: instead of calculating $n_i$ by a combination of processes with different activation rates, we regard the process between two communication events as almost entirely homogeneous and approximate the distribution of large $n_i$ by a geometric distribution with parameter
\begin{equation}
    \underline{\lambda}_j \simeq \min_{n'} \lambda_{j, n'}.
\end{equation}
In other words, we replace the activation-modulated process by a communication-modulated process. The distribution of ICANs is then given as 
\begin{equation}
    \begin{aligned}
    \phi(n_i) &= \int_0^\infty \underline{\lambda}_j(1 - \underline{\lambda}_j)^{n_i - 1} G(\underline{\lambda}_j)d\underline{\lambda}_j\\
    &\simeq \int_0^\epsilon \underline{\lambda}_j\exp(-\underline{\lambda}_j n_i) G(\underline{\lambda}_j)d\underline{\lambda}_j,
    \end{aligned}
    \label{eq:p_n}
\end{equation}
where $G(\underline{\lambda}_j)$ is the distribution of the activation probability for each ICT. The approximation in the second line is derived from the fact that the tail behavior of $\phi(n_i)$ will be dominated by the contributions from small $\underline{\lambda}_j$. We put a small finite cutoff $\epsilon$ and perform the integration up to this value. 

\begin{figure*}[tb]
    \centering
    \includegraphics[width=\textwidth]{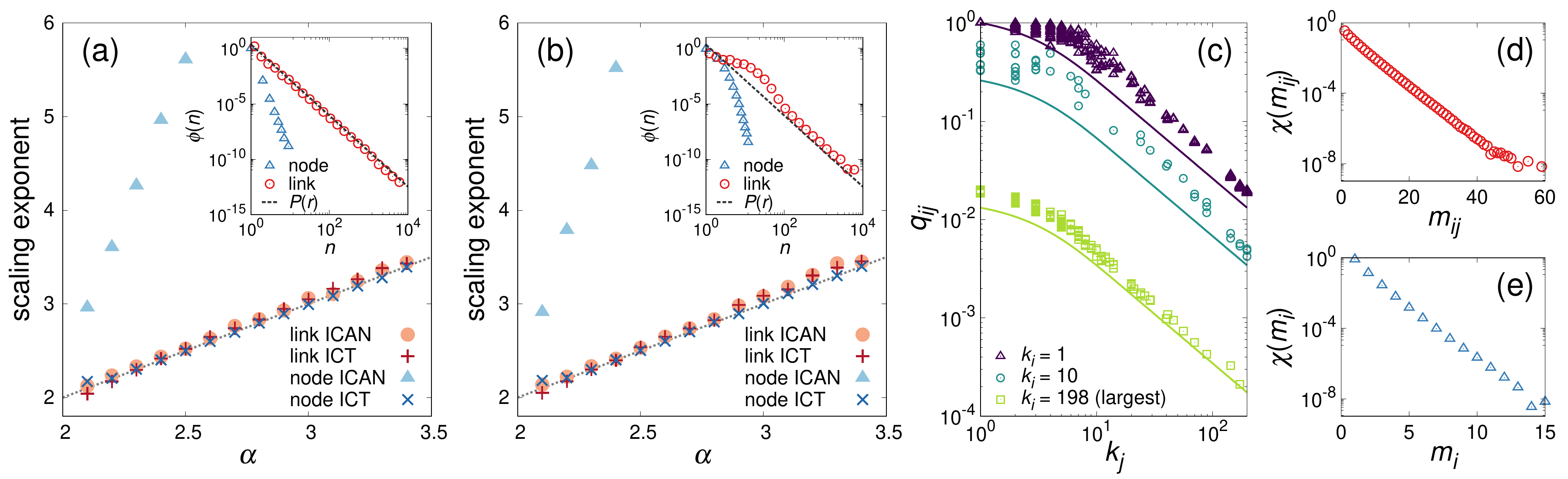}
    \caption{
    (a,~b) The scaling exponents of ICT and ICAN distributions, each for nodes and links, as a function of the scaling exponent $\alpha$ of IAT distributions in the cases of (a) the polyvalent and (b) monovalent models, respectively. The results are obtained from numerical simulations run for $T = 4 \times 10^7$ time steps on a random regular graph with $N = 100$ and $k = 6$. The dotted lines represent identity. The insets show the ICAN distributions when $\alpha = 3.2$. 
    (c) The probability $q_{ij}$ of communication between pairs of simultaneously activated adjacent nodes of degree $k_i$ and $k_j$ in a scale-free graph. The parameters are the same as in Fig.~\ref{fig:ICT_distributions}(i). The continuous lines, for each $k_i$, represent the theoretical curves given by Eq.~\eqref{eq:communication_prob_given_coactivation} with a numerically obtained value of $\rho$.
    (d, e) The distributions of (d) link ICCANs and (e) node ICCANs, both obtained from simulations of the monovalent model with $\alpha = 3.2$. The network topology is the same as in (b).}
    \label{fig:exponents}
\end{figure*}

In order to estimate the functional form of $G(\underline{\lambda})$, we go back to the original activation-modulated picture. For a given activation rate $\lambda$, the IATs are distributed as 
\begin{equation}
    p(r | \lambda) = \lambda e^{-\lambda r}.
\end{equation}
Conversely, the distribution of $\lambda$ conditional on $r$ is given by the Bayes' theorem,
\begin{equation}
    p(\lambda|r) = \frac{p(r|\lambda)p(\lambda)}{\int_0^\infty p(r|\lambda) p(\lambda) d\lambda}.
    \label{eq:Bayes}
\end{equation}
As we do not assume anything about the prior $p(\lambda)$ except $\lambda > 0$, we adopt the non-informative density, which is uniform throughout its domain. Eq.~\eqref{eq:Bayes} then reads
\begin{equation}
    p(\lambda | r) = r^2\lambda e^{-\lambda r}
\end{equation}
with the normalization factor $r^2$. When the IAT distribution scales as $P(r) \sim r^{-\alpha}$ at the tail, the following scaling holds for small values of $\lambda$:
\begin{equation}
    \begin{aligned}
    F(\lambda) &= \int_0^\infty p(\lambda | r) P(r) dr \\
    &\sim \lambda \int_0^\infty r^{2-\alpha} e^{-\lambda r}dr \sim \lambda^{\alpha - 2}.
    \end{aligned}
\end{equation}
This is consistent with the fact that for the event-modulated Poisson processes, $P(r)$ will have a power-law tail with exponent $\alpha$ when $F(\lambda)$ is a gamma distribution with shape parameter $\alpha - 1$~\cite{Masuda2018Gillespie}, which scales as $F(\lambda) \sim \lambda^{\alpha - 2}$ for small $\lambda$. We assume that $G(\underline{\lambda}) \simeq F(\underline{\lambda})$ for $0 < \underline{\lambda} \leq \epsilon$ and plug the scaling into Eq.~\eqref{eq:p_n} to obtain
\begin{equation}
    \phi(n_i) \sim \int_0^\epsilon \underline{\lambda}_j^{\alpha - 1} \exp(-\underline{\lambda}_j n_i)d\underline{\lambda}_j \sim {n_i}^{- \alpha}.
    \label{eq:ICAN_distribution_is_power_law}
\end{equation}
This derivation tells us that the statistics of the ICANs of node $i$ is determined by the activation process of node $j$. If the IAT distributions for node $i$ and $j$ are characterized by different exponents $\alpha_i$ and $\alpha_j$, respectively, then $\phi(n_i) \sim n_i^{- \alpha_j}$.

We now turn to our second question regarding the statistics of ICTs as the sum of an ICAN of IATs. Since the ICAN and IAT are independent random variables, we exploit the analytical results in Ref.~\cite{Jo2013Contextual}: We consider the following sum 
\begin{equation}
    \tau = \sum_{n' = 1}^n r_{n'}
\end{equation}
where the summands $r$ and the number of summands $n$ are independent random variables and they both follow power-law distributions, $P(r) \sim r^{-\alpha}$ and $\phi(n) \sim n^{-\beta}$. Then, $\tau$ also asymptotically obeys a power-law distribution $\psi(\tau) \sim \tau^{-\alpha'}$ where
\begin{equation}
    \alpha' = \min\{(\alpha - 1)(\beta - 1) + 1, \alpha, \beta\}.
    \label{eq:exponent_rel}
\end{equation}
In our case, since the IAT $r_i$ and ICAN $n_i$ in Eq.~\eqref{eq:ICT_is_equal_to_sum_of_IAT} are shown to have the same scaling exponent as $\beta = \alpha$, the ICT distribution also follows a power law with the same exponent $\alpha' = \alpha$, that is,  
\begin{equation}
    \psi(\tau) \sim \tau^{-\alpha}.
\end{equation}

\subsection{Polyvalent model}

The polyvalent model assumes that communication occurs on a link every time when the two end nodes are activated at the same time, irrespective of the states of other nodes in the system. Therefore, the dimer case discussed in the previous subsection directly translates into the communication patterns on links (compare the left panel of (a) to (b) in Fig.~\ref{fig:schematic2}). Indeed, Fig.~\ref{fig:exponents}(a) shows that the link ICAN distributions follow power laws, and that the scaling given by Eq.~\eqref{eq:ICAN_distribution_is_power_law} agrees well with the numerical results. The distribution of polyvalent link ICTs is the same as that of ICTs for the dimer case and given by 
\begin{equation}
    \psi_\mathrm{p}(\tau^\mathrm{p}_{ij}) \sim \left[\tau^\mathrm{p}_{ij}\right]^{-\alpha}.
\end{equation}

In order to investigate the node communication processes, we introduce a time frame defined by counting the number of activations of a node, formally expressed as
\begin{equation}
    \nu_i(t) = \sum_{l = 0}^{\infty}\theta\left(t - t_{i, l}\right),
    \label{eq:activation_clock}
\end{equation}
where $t$ is the wall-clock time, $t_i$ are the times that node $i$ is activated, and $\theta(\cdot)$ is the Heaviside step function.
Simply put, the activation-based time $\nu_i$ is measured by a clock that ticks one unit forward upon every activation of node $i$. This time transformation $t \mapsto \nu_i(t)$ rescales an ICT $\tau = t' - t''$ into an ICAN $n_i = \nu_i(t') - \nu_i(t'')$, meaning that an ICAN is an ``inter-communication time" for the processes in the time frame $\nu_i$. We note that similar concepts of time frame transformation, named ``relative clock" and ``activity clock," are used in recent studies~\cite{Zhou2012Relative, Panisson2012Dynamics, Gauvin2013Activity}.

In the wall-clock time frame, the communication processes on adjacent links $ij$ and $ij'$ are correlated because of the underlying activation process of node $i$. However, in the activation-based time frame $\nu_i$, in which the activations of node $i$ are regularized, the two communication processes are independent because when $i$ is activated, communication between nodes $i$ and $j$ depends only on whether $j$ is simultaneously activated and is not affected by the behavior of node $j'$. Since the communication process on each link is characterized by power-law distributed ICANs as in Eq.~\eqref{eq:ICAN_distribution_is_power_law}, the node communication process as the superposition of independent link processes has a thinner-tailed ICAN distribution, that is, 
\begin{equation}
    \phi_\mathrm{p}\left(n^\mathrm{p}_i\right) \sim \left[n^\mathrm{p}_i\right]^{-\beta}
\end{equation}
with $\beta > \alpha$ (see Fig.~\ref{fig:exponents}(a)). 

By taking the same approach as in the previous subsection, we write a node ICT as follows:
\begin{equation}
    \tau^\mathrm{p}_i = \sum_{n' = 1}^{n^\mathrm{p}_i} r_{n'},
\end{equation}
where $P(r) \sim r^{-\alpha}$. Then, by following the scaling relation given by Eq.~\eqref{eq:exponent_rel}, we find that the polyvalent node ICTs are distributed as 
\begin{equation}
    \psi_\mathrm{p}\left(\tau^\mathrm{p}_i\right) \sim \left[\tau^\mathrm{p}_i\right]^{-\alpha}.
\end{equation}

\subsection{Monovalent model}

In contrast to the polyvalent model, simultaneous activation of two adjacent nodes do not necessarily trigger communication between them in the monovalent model. In order to study the statistics of link ICTs, we first need to account for the random pattern of successful communication when a pair of adjacent nodes are simultaneously activated.

Suppose that node $i$ with degree $k_i$ is activated at a time step along with $\kappa_i + 1$ activated neighbors including node $j$. Because activation processes of different nodes are independent of each other, variable $\kappa_i$ is binomially distributed as 
\begin{equation}
    \kappa_i \sim B(\kappa_i; k_i - 1, \rho) = \binom{k_i - 1}{\kappa_i} \rho^{\kappa_i} (1 - \rho)^{(k_i - 1) - \kappa_i},
\end{equation}
where $\rho = 1 / \langle r \rangle$ denotes the probability that each neighbor of node $i$ is activated when node $i$ is activated. If $\kappa_i > 0$, the communication between nodes $i$ and $j$ occurs only if $i$ selects $j$ as the counterpart as a result of random matching. Although the probability of selecting each of the activated neighbors is not uniform in general, we assume the uniformity for simplicity so that the probability that node $j$ is selected is equal to $1 / (\kappa_i + 1)$. The same goes for node $j$ for its $\kappa_j + 1$ activated neighbors including node $i$. Then, the probability that simultaneous activation of nodes $i$ and $j$ leads to communication between them is approximated by
\begin{equation}
    \begin{aligned}
    q_{ij} &= \sum_{\kappa_i = 0}^{k_i - 1}\frac{B(\kappa_i; k_i - 1, \rho)}{\kappa_i + 1}\sum_{\kappa_j = 0}^{k_j - 1}\frac{B(\kappa_j; k_j - 1, \rho)}{\kappa_j + 1}\\
    &= \frac{\left[1 - (1 - \rho)^{k_i}\right]\left[1 - (1 - \rho)^{k_j}\right]}{\rho^2 k_i k_j}.
    \end{aligned}
    \label{eq:communication_prob_given_coactivation}
\end{equation}
This form reduces to $q_{ij} = 1$ for the dimer case of $k_i = k_j = 1$, in which they communicate with each other every time they are simultaneously activated. In the limit where $k_i, k_j \gg 1$, we have $q_{ij} \simeq 1 / \rho^2 k_i k_j$. Equation~\eqref{eq:communication_prob_given_coactivation} is, on the whole, numerically supported as shown in Fig.~\ref{fig:exponents}(c), although deviations and fluctuations are notable. We think these deviations originate from perturbation by higher-order effects involving more than two nodes, which violates the uniformity assumption.

Let $m_{ij}$ be the number of times that adjacent nodes $i$ and $j$ are simultaneously activated between two consecutive communication events, including their activation that triggered the latter of the two communication events but excluding the one that triggered the former (see Fig.~\ref{fig:schematic2}(c)). We call $m_{ij}$ an inter-communication coactivation number (ICCAN) of link $ij$. Because random pairing is done independently at each time step, $m_{ij}$ is geometrically distributed (see Fig.~\ref{fig:exponents}(d)) as
\begin{equation}
    \chi(m_{ij}) = (1 - q_{ij})^{m_{ij} - 1} q_{ij}.
    \label{eq:ICCAN_distribution_is_geometric}
\end{equation}

As depicted in Fig.~\ref{fig:schematic2}(c), a monovalent link ICT, denoted by $\tau^\mathrm{m}_{ij}$, is equal to the sum of $m_{ij}$ successive polyvalent link ICTs, 
\begin{equation}
    \tau^\mathrm{m}_{ij} = \sum_{m' = 1}^{m_{ij}} \tau^\mathrm{p}_{ij, m'}.
    \label{eq:monovalent_link_ICT_is_equal_to_sum_of_polyvalent_link_ICT}
\end{equation}
The distribution of $\tau^\mathrm{m}_{ij}$ is written as
\begin{equation}
    \psi_\mathrm{m}(\tau^\mathrm{m}_{ij}) = \sum_{m_{ij}=1}^\infty h(\tau^\mathrm{m}_{ij}; m_{ij}) \chi(m_{ij}),
\end{equation}
where
\begin{equation}
    \begin{aligned}
    h(\tau; m) \equiv &\sum_{\tau_1 = 0}^{\infty} \dots \sum_{\tau_m = 0}^{\infty} \psi_\mathrm{p}(\tau_1) \dots \psi_\mathrm{p}(\tau_m) \\
    &\times \delta\left(\tau - \sum_{m'=1}^m \tau_{m'}\right)
    \end{aligned}
\end{equation}
is the probability that a monovalent link ICT is equal to $\tau$ and it is composed of $m$ polyvalent link ICTs. Here $\delta(\cdot)$ denotes the Dirac delta function. 

An analytical evaluation of the discrete power-law distribution $\psi_\mathrm{p}(\tau)$ is not straightforward. Instead, we consider a continuous counterpart given by
\begin{equation}
    \psi_\mathrm{p}(\tau^\mathrm{p}_{ij}) = (\alpha - 1) \left[\tau^\mathrm{p}_{ij}\right]^{-\alpha}\theta(\tau^\mathrm{p}_{ij} - 1),
    \label{eq:continuous_polyvalent_link_ICT_distribution}
\end{equation}
where $\alpha > 1$. The Laplace transform of Eq.~\eqref{eq:continuous_polyvalent_link_ICT_distribution} is given as  
\begin{equation}
    \tilde{\psi}_\mathrm{p}(s)=(\alpha-1)s^{\alpha-1}\Gamma(1-\alpha,s),
\end{equation}
where $\Gamma(\cdot, \cdot)$ denotes the upper incomplete gamma function.
In the asymptotic limit of $s\to 0$, one gets
\begin{equation}
    \tilde{\psi}_\mathrm{p}(s) = 1 + bs^{\alpha-1} + cs + \mathcal{O}(s^2)
\end{equation}
with $b \equiv \Gamma(1-\alpha)(\alpha-1)$, where $\Gamma(\cdot)$ is the gamma function, and $c\equiv (\alpha-1)/(2-\alpha)$. By only keeping the leading terms of expansion with respect to $s$, we have 
\begin{equation}
    \tilde{\psi}_\mathrm{m}(s) \simeq 1 + \frac{bs^{\alpha - 1} + cs}{q_{ij}}.
    \label{eq:Laplace_transform_of_monovalent_link_ICT_distribution}
\end{equation}
The inverse Laplace transform of Eq.~\eqref{eq:Laplace_transform_of_monovalent_link_ICT_distribution} in the limit of $\tau \to \infty$ yields
\begin{equation}
    \psi_\mathrm{m}(\tau^\mathrm{m}_{ij}) \simeq \frac{\alpha - 1}{q_{ij}}\left[\tau^\mathrm{m}_{ij}\right]^{-\alpha}.
    \label{eq:monovalent_link_ICT_follows_power_law}
\end{equation}
Equation~\eqref{eq:monovalent_link_ICT_follows_power_law} is valid for any values of $\alpha$ because considering higher-order terms in the expansion of Eq.~\eqref{eq:Laplace_transform_of_monovalent_link_ICT_distribution} does not affect the asymptotic form given by Eq.~\eqref{eq:monovalent_link_ICT_follows_power_law}. This result indicates that the link ICT distribution for the monovalent model has a power-law tail with the same exponent as the link ICT distribution for the polyvalent model, which is consistent with the numerical results presented in Fig.~\ref{fig:exponents}(b). At the same time, the geometric distribution of ICCANs contributes to the hump at the bulk part of the monovalent link ICT distribution. For a dense network where the degrees are generally large, $q_{ij}$ is small and the geometric decay of $\chi$ in Eq.~\eqref{eq:ICCAN_distribution_is_geometric} is slow; this is why the hump is larger in denser networks as shown in Fig.~\ref{fig:ICT_distributions}.

Lastly, we discuss the node ICT distribution for the monovalent model. Unlike the polyvalent case, the monovalent communication events on adjacent links (i.e., links sharing a node) are not independent of each other even in the activation-based time frame because communication between a pair of nodes forbids the nodes to communicate with other nodes at the same time. Nevertheless, Fig.~\ref{fig:exponents}(e) shows that node ICCANs $m_i$, i.e., the numbers of times that node $i$ is activated simultaneously with any of its neighbors until it communicates with one of them, are geometrically distributed. This observation suggests that the probability that a node succeeds to communicate with another node is constant every time it is simultaneously activated with at least one of its neighbors. 

A monovalent node ICT, denoted by $\tau^\mathrm{m}_i$, can be written as the sum of $m_i$ successive polyvalent node ICTs as follows:
\begin{equation}
    \tau^\mathrm{m}_i = \sum_{m' = 1}^{m_i} \tau^\mathrm{p}_{i,m'}.
    \label{eq:monovalent_node_ICT_is_equal_to_sum_of_polyvalent_node_ICTs}
\end{equation}
Equation~\eqref{eq:monovalent_node_ICT_is_equal_to_sum_of_polyvalent_node_ICTs} is analogous to the relation between the monovalent and polyvalent link ICTs given by Eq.~\eqref{eq:monovalent_link_ICT_is_equal_to_sum_of_polyvalent_link_ICT}. Again using the scaling relation given by Eq.~\eqref{eq:monovalent_link_ICT_follows_power_law}, one obtains the monovalent node ICT distribution with a power law at its tail as follows:
\begin{equation}
    \psi_\mathrm{m} \left(\tau^\mathrm{m}_i\right) \sim \left[\tau^\mathrm{m}_i\right]^{-\alpha}.
\end{equation}
This result is in good agreement with the numerically obtained scaling relations shown in Fig.~\ref{fig:exponents}(b). 

\section{Conclusion}
\label{sec:conclusion}
In order to explain the origin of the bursty activity patterns of nodes and links observed in empirical communication systems, we have proposed a temporal network model where the nodes communicate with each other according to their non-Poissonian random activation. The two variants of the model that we discussed are both able to reproduce heavy tails in the inter-communication time (ICT) distributions for nodes and links for various network topologies. We have shown that the polyvalent ICTs are power-law distributed because each of them is a sum of inter-activation times (IATs) where both the summands and the number of summands are power-law distributed random numbers, which stem from the node activation processes. A monovalent ICT is described as a sum of polyvalent ICTs, where the number of summands is geometrically distributed because the activated nodes are paired randomly and independently at each time step. As a result, an exponential hump appears in the bulk part of the ICT distributions, especially prominently for small-degree nodes and for links between large-degree nodes; nevertheless, the tail part of the distributions follows a power law with the same scaling exponent as the power law in the IAT distribution.

The superposition of independent communication sequences with power-law distributed ICTs does not yield a sequence with the ICTs distributed as a power law with the same scaling exponent. Therefore, the assumption of independence between links is unable to account for the real-world observations. Our results suggest a possible mechanism behind the reconciliation between bursty dynamics of nodes and of links: Link-link correlations emerge as a result of underlying node activations, each of which may or may not realize actual communication.

Further steps can be taken in this line of research. 
In this work, we have considered a homogeneous population of nodes that shares the same activation statistics. In reality, the activity levels of nodes and the weights of links, i.e., the frequency of communication between pairs of individuals, are heterogeneous~\cite{Onnela2007Analysis}. It would be straightforward to include such heterogeneity into our modeling framework if we consider nodes endowed with different scaling exponents of the IAT distribution. 
We have also assumed that a node behaves in an equal manner toward every neighbor. However, empirical data show that individuals allocate their efforts to communicate with others unevenly among alters~\cite{Saramaki2014Persistence}. This effect can be taken into account in the monovalent model by setting biases toward certain links when pairing communication partners. 
Another possible extension is to implement communication among a group of nodes, which corresponds to ``conference calls" or ``group chats," in a direction similar to Ref.~\cite{Petri2018Simplicial}. 
One can also tailor the temporal structure of node activation patterns to account for the empirical observation of long-range correlated node ICTs in human communication~\cite{Karsai2012Universal, Jo2019Bursttree}.
Future work also includes how the presence of link-link correlations affects dynamical processes taking place in temporal networks and the associated network control problems~\cite{Li2017Fundamental}.

\begin{acknowledgments}
    T.H. thanks M.I.D. Fudolig for careful reading of the manuscript. A.L. was supported by the International Human Frontier Science Program (HFSP) Postdoctoral Fellowship (Grant No. LT000696/2018-C) and Foster Lab at Oxford. H.-H.J. was supported by Basic Science Research Program through the National Research Foundation of Korea (NRF) funded by the Ministry of Education (NRF-2018R1D1A1A09081919).
\end{acknowledgments}


%

\end{document}